\documentclass[12pt]{iopart}
\usepackage{graphicx}
\usepackage{setstack,cite}
\begin{document}
\date{}

\title[General solution for the modified Emden type equation]
{On the general solution for the modified 
Emden type equation $\ddot{x}+\alpha x\dot{x}+\beta x^3=0$}

\author{V~K~Chandrasekar, M~Senthilvelan and M~Lakshmanan}

\address{Centre for Nonlinear Dynamics, Department of Physics,  Bharathidasan
University, Tiruchirapalli - 620 024, India.}

\begin{abstract} 
In this paper, we demonstrate 
that the modified Emden type equation (MEE), 
$\ddot{x}+\alpha x\dot{x}+\beta x^3=0$, is integrable either explicitly or by
quadrature for any value of $\alpha$ and  $\beta$. We also prove that the MEE
possesses appropriate time-independent Hamiltonian function for the full range 
of parameters $\alpha$ and $\beta$. In addition, we show that the MEE is
intimately connected with two well known nonlinear models, namely the
force-free Duffing type  oscillator equation and the two dimensional
Lotka-Volterra (LV) equation and  thus the complete integrability of the latter
two models can also be understood in terms of the MEE.
\end{abstract}
\section{Introduction}
\label{sec1}
One of the well discussed models in nonlinear dynamics is the modified Emden  
type equation (MEE), also called the modified Painlev\'e-Ince equation, 
\begin{eqnarray} 
\ddot{x}+\alpha x\dot{x}+\beta x^3=0,
 \label {eq01}
\end{eqnarray}
where  over dot denotes differentiation with respect to time and 
$\alpha$ and $\beta$ are arbitrary parameters. This equation has
received attention from both mathematicians and physicists for more than a 
century \cite{Painleve,Ince,Davis,
Kamke,Murphi}. For example, in the nineteenth 
century, Painlev\'e had 
studied this equation and identified general solution for two parametric 
choices, namely, $(i)\;\beta=\frac{\alpha^2}{9}$ and 
$(ii)\;\beta=-\alpha^2$ \cite{Painleve,Ince}. The above
differential equation (\ref{eq01}) arises in a variety of mathematical 
problems such as univalued 
functions defined by second order differential equations \cite{Gobulev} and
the Riccati equation \cite{Chisholm}. On the other hand
physicists have shown that this equation arises in different contexts: 
for example, it occurs in the study of 
equilibrium configurations of a spherical gas cloud acting under the mutual 
attraction of its molecules and subject to the laws of thermodynamics 
\cite{Dix:1990,Moreira} and in the modelling of the fusion of 
pellets \cite{Erwin}. It also governs spherically symmetric expansion or
collapse of a relativistically gravitating mass \cite{McVittie}. This equation
can also be thought of as a one-dimensional analogue \cite{Chisholm,Yang} of 
the boson `gauge-theory'
equations introduced by Yang and Mills. 
Apart from the above, for the past two decades or so the invariance and integrability
properties of this equation alone have been studied in detail by a number of 
authors,
see for example Ref. \cite{mahomed:1985,Duarte2,Bouquet,Sarleti,leach:1988a,
Steeb,feix:1997,Ibragimov1,Leach1,Chand,Chand1}. In a nutshell, the MEE 
(\ref{eq01}) 
has been found to possess explicit general solution only for the following 
parametric choices, that is, 
$(i)\; \alpha=0,\;(ii)\;\beta=0,\;(iii)\;\beta=\frac{\alpha^2}{9}$ and 
$(iv)\;\beta=-\alpha^2$. While the cases (i) and (ii) can be integrated
trivially, in the third case the equation is linearizable to a free particle
equation and in the fourth case the general solution can be expressed in terms
of Weierstrass elliptic function \cite{Painleve,Ince,Davis,
Kamke,Murphi,mahomed:1985,Duarte2,Bouquet,Sarleti,leach:1988a,Steeb,feix:1997,
Ibragimov1,Leach1,Chand,Chand1,Euler}. 
Equation (\ref{eq01}) has also been noted to possess Painlev\'e property only 
for certain values of $r=\frac{\alpha}{4\beta}(\alpha\pm\sqrt{\alpha^2-8\beta})$ 
\cite{leach:1988a,feix:1997}. Finally, we mention that equation
(\ref{eq01}) admits a two parameter Lie point symmetry group for arbitrary
values of $\alpha$ and $\beta$, while for the chioce $\beta=\frac{\alpha^2}{9}$
the equation (\ref{eq01}) possesses eight-parameter Lie point symmetries
\cite{mahomed:1985}. However, the 
general solution for the equation (\ref{eq01}) with $\alpha$ and $\beta$ are 
arbitrary is yet to be explored.

Very recently \cite{Chand1}, the present authors have studied a generalized
version of this equation from a different perspective and shown that the equation
(\ref{eq01}), for $\alpha^2 \geq 8\beta$, possesses time independent integrals 
and admits a Hamiltonian formalism which in turn ensures its complete 
integrability \cite{Chand1}. However, due to the complicated form of
the first integral the general solution was not obtained in the previous work. 
Keeping in mind the historical importance and popularity of this model we 
kept exploring the general solution of equation (\ref{eq01}) in phase-space. 
Based on our investigation, in this paper, we construct the time 
independent integrals for the equation (\ref{eq01}) for arbitrary values of 
$\alpha$ and $\beta$ (including the case $\alpha^2 < 8\beta$). Since the first 
integrals are not in simple polynomial forms it
is difficult to obtain the general solution by just directly integrating them. 
In order to overcome this difficulty,  first we identify time 
independent Hamiltonians from these time independent integrals
and making use of suitable canonical transformations we convert the
Hamiltonians into standard forms. We then integrate the new Hamiltonians and
obtain the general solutions or reduce to quadratures. In this way 
we report the general solution for the equation (\ref{eq01}) for 
arbitrary values of $\alpha$ and $\beta$ for the first time. Our motivation 
to explore this
solution is, as we see below, also due to the fact that the MEE (\ref{eq01}) 
is not a stand-alone
model. It is intimately connected to force-free Duffing oscillator type equation 
(vide equation (\ref{eq08}) below) and two dimensional Lotka-Volterra (LV) equation 
(vide equation
(\ref{eq10}) below). The popularity and importance of these two models need no
emphasis \cite{Chand1,Bluman,Laks1,Murray,Cairo1,Cairo2}. Thus 
exploring the general solution for the 
equation (\ref{eq01}) also serves to establish the complete integrability 
of these two models for appropriate parameters besides understanding the 
dynamics of the other models mentioned in the introduction.

The plan of the paper is as follows. In Section \ref{sec2}, we give the 
time independent integrals and corresponting Hamiltonians for the MEE 
(\ref{eq01}). Using suitable canonical transformations, we obtain the 
general solutions for the parameter ranges $\alpha^2=8\beta,
\;\alpha^2 >8\beta$ and $\alpha^2 <8\beta$ separately in Section \ref{sec3}. In Section 
\ref{sec4}, 
we show that the MEE (\ref{eq01}) is intimately connected with two other 
well known nonlinear models, namely, the force-free Duffing oscillator type
equation and two dimensional Lotka-Volterra equation. 
Finally, in Section \ref{sec5} we summarize our 
results.

\section{Time independent integrals and Hamiltonian description}
\label{sec2}
As pointed out in the introduction, the MEE (\ref{eq01}) cannot be
straightforwardly integrated. Making use of the modified Prelle-Singer method
developed by us recently \cite{Chand,Chand1} following the earlier work of 
Duarte et al
\cite{Duarte}, in this section we first identify the first integrals 
separately for each of
the three ranges (i) $\alpha^2 =8\beta$, 
(ii) $\alpha^2 >8\beta$ and (iii) $\alpha^2 <8\beta$. Then we identify suitable
canonical Hamiltonian description for each of these cases.
\subsection{Time independent integrals}
\label{sec20}
In Ref. \cite{Chand1}, the time independent integrals for the equation 
(\ref{eq01}) with $\alpha^2 \geq 8\beta$ has been reported using the modified 
Prelle-Singer procedure. However, improving the ansatz
given in Ref. \cite{Chand1} one can obtain the time independent integrals
for all values of $\alpha$ and $\beta$ and the method of deriving the 
integrals for the equation (\ref{eq01}) with $\alpha$ and $\beta$ arbitrary 
is given in Appendix A. Using this method we identify the following  
time independent first integrals for the 
equation (\ref{eq01}) with $\alpha$ and $\beta$ arbitrary, that is,
\begin{eqnarray}
\fl\mbox{Case 1:}\;\alpha^2=8\beta &\;\;(r=2)\nonumber\\ 
&I=\log(\alpha^2x^2+4\alpha \dot{x})
-\frac{4\alpha\dot{x}}{\alpha ^2x^2+4\alpha\dot{x}}&
\label {intm03}\\
\fl\mbox{Case 2:}\; \alpha^2 >8\beta &\;\;(r\neq0,1,2) \nonumber\\
&I=\frac{(r-1)}{(r-2)}
\bigg(\dot{x}+\frac{(r-1)}{2r}\alpha x^2\bigg)^{1-r}
 \bigg(\dot{x}+\frac{\alpha }{2r}x^2\bigg)
& \label {intm02}\\
\fl\mbox{Case 3:}\; \alpha^2 <8\beta & \nonumber\\ 
&I=\frac{1}{2}\log(2\dot{x}^2+\alpha x^2\dot{x}+\beta x^4)
+\frac{\alpha}{\omega}tan^{-1}\bigg[\frac{\alpha \dot{x}+2\beta x^2}
{\omega \dot{x}}\bigg],& \label {intm01}
\end{eqnarray}
where $\omega=\sqrt{8\beta-\alpha^2}$ and 
$r=\frac{\alpha}{4\beta}(\alpha\pm\sqrt{\alpha^2-8\beta})$.  
Note that $r=0$ and $1$ correspond to the trivial cases $\alpha=0$ and
$\beta=0$, respectively, and so they are not considered here separately. 

As it is very difficult to integrate equations (\ref{intm03})-(\ref{intm01}) and 
obtain the general solutions by direct integration, we
correlate these integrals with appropriate Hamiltonians. In the following we
briefly give the method of obtain the Hamiltonian from the known time 
independent integral.
\subsection{Hamiltonian description}
\label{sec21}
To explore the Hamiltonian description of (\ref{intm03}), let us assume the 
existence of a Hamiltonian
\begin{eqnarray}
I(x,\dot{x}) = H(x,p) = p \dot{x}-L(x,\dot{x}),
\label{mlin13}
\end{eqnarray}
where $L$ is the Lagrangian and $p$ is the canonically conjugate
momentum. Then
\begin{eqnarray}
&&\frac{\partial{I}}{\partial\dot{x}} = \frac{\partial{H}}{\partial\dot{x}} = 
\frac{\partial{p}}{\partial\dot{x}}\dot{x} + p - \frac{\partial{L}}{\partial\dot{x}} = 
\frac{\partial{p}}{\partial\dot{x}} \dot{x}. 
\label{mlin13a}
\end{eqnarray}
From (\ref{mlin13a}) we identify
\begin{eqnarray}
p =\int \frac{I_{\dot{x}}}{\dot{x}} d\dot{x}+f(x),
\label{mlin13b}
\end{eqnarray}
where $f(x)$ is an arbitrary function of $x$ and the Lagrangian $L=p\dot{x}
-I(x,\dot{x})$.
Here, without loss of generality, we take $f(x)=0$. Substituting the integrals 
(\ref{intm03})-(\ref{intm01}) into (\ref{mlin13b}) and integrating 
the resultant integrals we can obtain the expression for the canonical 
momentum $p$. Substituting 
back the latter into the equation~(\ref{mlin13}) and simplifying the resultant 
equation we arrive at the following Lagrangian, 
\begin{eqnarray}
L=\left\{
\begin{array}{ll}
-\log(\dot{x}+\frac{\alpha}{4} x^2), & \;\;\; \alpha^2=8\beta\\
\frac{1}{(2-r)}(\dot{x}+\frac{(r-1)}{2r}\alpha x^2)^{(2-r)},
&\;\;\;\alpha^2>8\beta\\
\frac{1}{\omega}
\bigg(tan^{-1}\bigg[\frac{4\dot{x}+\alpha x^2}{\omega x^2}\bigg]
(\frac{4\dot{x}}{x})-\alpha tan^{-1}\bigg[\frac{\alpha \dot{x}+2\beta x^2}
{\omega \dot{x}}\bigg]\bigg)\\
\qquad \qquad -\frac{1}{2}\log(2\dot{x}^2+\alpha x^2\dot{x}+\beta x^4),
&\;\;\;\alpha^2 <8\beta 
\end{array}\right.\label{mlin14}
\end{eqnarray}
and Hamiltonian
\begin{eqnarray}
H=\left\{
\begin{array}{ll}
\log(-\frac{4\alpha}{p})-\frac{\alpha}{4}px^2,
&\qquad\qquad\qquad\alpha^2=8\beta \\
\displaystyle{\frac{(r-1)}{(r-2)}(p)^{\frac{r-2}{r-1}}
-\frac{(r-1)}{2r}\alpha x^2p},&\qquad\qquad\qquad
\alpha^2>8\beta \\
\displaystyle{\frac{1}{2}\log[x^4 sec^2(\frac{\omega}{4} x^2p)]
-\frac{\alpha }{4}x^2p},
&\qquad\qquad\qquad\alpha^2 <4\beta  
\end{array}\right.\label{mlin16}
\end{eqnarray}
where the canonically conjugate momentum 
\begin{eqnarray}
p=\left\{
\begin{array}{ll}
-\frac{1}{(\dot{x}+\frac{\alpha}{4}x^2)}
&\qquad\qquad\qquad\qquad\quad\;\; \alpha^2=8\beta\\
\displaystyle{\bigg(\dot{x}+\frac{(r-1)}{2r}\alpha x^2
\bigg)^{1-r}}&\qquad\qquad\qquad\qquad\quad\;\;\alpha^2>8\beta\\
\displaystyle{\frac{4}{\omega x^2}
tan^{-1}\bigg[\frac{4\dot{x}+\alpha x^2}{2\omega x^2}\bigg]},
&\qquad\qquad\qquad\qquad\quad\;\; \alpha^2 <8\beta
\end{array}\right.\label{mlin15}
\end{eqnarray}
respectively for the MEE (\ref{eq01}).

One can easily check that the canonical equations of motion for the above
Hamiltonians are nothing but the equation of motion (\ref{eq01}) of the 
MEE in the appropriate parametric regimes. 

\section{General solutions}
\label{sec3}
In this section, we consider each of the above three cases separately, 
namely, (i) $\alpha^2 =8\beta$, 
(ii) $\alpha^2 >8\beta$ and (iii) $\alpha^2 <8\beta$,  
and obtain their respective general solutions using suitable canonical 
transformations.
\subsection{Case 1: $\alpha^2=8\beta$}
\label{sec22}
To derive the general solution for this parametric choice first we consider the
Hamiltonian given in (\ref{mlin16}) for $\alpha^2=8\beta$ as 
\begin{eqnarray}
H=\log(-\frac{4\alpha}{p})-\frac{\alpha}{4}px^2.\label{eq04}
\end{eqnarray}
Introducing the canonical transformation 
\begin{eqnarray}
x=\frac{4P}{\alpha U},\qquad\;\;p=-\frac{\alpha U^2}{8},
\label {eq05}
\end{eqnarray}
the Hamiltonian (\ref{eq04}) can be recast into the standard form
\begin{eqnarray}
H=\frac{1}{2}P^2+\log(\frac{32}{U^2})\equiv \hat{E},\label{eq06a}
\end{eqnarray} 
where $\hat{E}$ is a constant. From the corresponding canonical equations 
$\dot{U}=P,\;\dot{P}=\frac{2}{U}$,
equation (\ref{eq06a}) can be rewritten as 
\begin{eqnarray}
E=\frac{1}{2}\dot{U}^2-2\log(U),\;\;\; E=\hat{E}-\log(32).\label{eq06}
\end{eqnarray}
Rewriting the above equation we get
\begin{eqnarray}
\frac{dU}{\sqrt{2E+4\log(U)}}=dt.\label{eq06b}
\end{eqnarray}
Integrating the equation (\ref{eq06b}) we obtain
\begin{eqnarray} 
U =\displaystyle{e^{-\frac{1}{2}\bigg(E+2\mbox{erf}^{-1}(z)^2\bigg)}},
\label{eq07a}
\end{eqnarray}
where $\displaystyle{z=\frac{2e^{\frac{E}{2}}(t_0+it)}{\sqrt{\pi}}}$ and 
$t_0$ is the second arbitrary integration constant and {\it erf} is the error 
function 
\cite{Abramowitz}.

Substituting the equation (\ref{eq07a}) into equation (\ref{eq05})
we get the general solution for the equation (\ref{eq01}), with the 
parametric choice $\beta=\frac{\alpha^2}{8}$, in the form (after some
modifications)
\begin{eqnarray} 
x(t) =\displaystyle{\frac{8}{\alpha}\mbox{erf}^{-1}(\bar{z})
e^{\frac{1}{2}\bigg(E+2\mbox{erf}^{-1}(\bar{z})^2\bigg)}}, \;\;
\bar{z}=\frac{2e^{\frac{E}{2}}(t_0-t)}{\sqrt{\pi}}.
\label{eq07}
\end{eqnarray}
In Fig. 1, we have plotted the solution of MEE (\ref{eq01}) given by the
expression (\ref{eq07}) for the parametric choice $\alpha^2=8\beta$ for
different initial conditions.
\begin{figure}[!ht]
\begin{center}
\includegraphics[width=.5\linewidth]{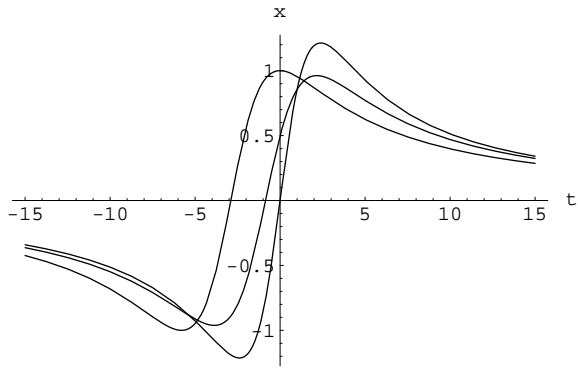}
\caption{Solution plot for case $\alpha^2=8\beta$ for different 
initial conditions.}
\end{center}
\end{figure}

\subsection{Case 2: $\alpha^2>8\beta$}
\label{sec23}
Now let us consider the Hamiltonian for the case $\alpha^2>8\beta$ from 
(\ref{mlin16}), that is,
\begin{eqnarray}
H=\displaystyle{\frac{(r-1)}{(r-2)}(p)^{\frac{r-2}{r-1}}
-\frac{(r-1)}{2r}\alpha x^2p}.\label{com03}
\end{eqnarray}

Interestingly, here we identify a canonical transformation for the 
Hamiltonian (\ref{com03}) in the form
\begin{eqnarray}
&x=-a\frac{P}{U},\qquad p=\frac{ U^2}{2a},
\label {com05}
\end{eqnarray}
where $a=\frac{2r}{\alpha(1-r)}$. It is straightforward to check that when $U$ and 
$P$ are canonical so do $x$ 
and $p$ (and vice versa) so that the Hamiltonian $H$ in Eq.~(\ref{com03}) can 
be rewritten as 
\begin{eqnarray}
&H=\frac{1}{2}P^2-k U^{m}\equiv E,
\label{com06}
\end{eqnarray} 
where $\displaystyle{m=\frac{2(r-2)}{(r-1)}}$ and 
$\displaystyle{k=\frac{(1-r)}{(2a)^{\frac{m}{2}} (r-2)}}$. From the canonical 
equations then we have $\dot{U}=P$ and $\dot{P}=mkU^{m-1}$.

To obtain the general solution, 
we introduce a transformation $X=U^m$ in equation (\ref{com06}) 
so that the latter can be brought to a quadrature of the form 
\begin{eqnarray}
t-t_0=\int \frac{X^{\frac{1-m}{m}}dX}{m\sqrt{2E+2kX}}, \;\; m\neq0,2,4.
\label{int01}
\end{eqnarray}
Now fixing $\frac{1-m}{m}=n$, the above integral leads us to the following
expression  \cite{Gradshteyn}, that is,
\begin{eqnarray}
\fl t-t_0=\int \frac{X^{n}dX}{m\sqrt{2E+2kX}}
=\frac{1}{m}\bigg(\frac{X^n\sqrt{2E+2kX}}{2k(n+\frac{1}{2})}
-\frac{2En}{2k(n+\frac{1}{2})}\int \frac{X^{n-1}dX}{\sqrt{2E+2kX}}\bigg).
\label{int02}
\end{eqnarray}
On the other hand fixing $\frac{1-m}{m}=-n$ in (\ref{int01}), we end up with the
following expression \cite{Gradshteyn}
\begin{eqnarray}
\fl \qquad\qquad  t-t_0&=&\int \frac{dX}{mX^{n}\sqrt{2E+2kX}}
\nonumber\\
&=&\frac{1}{m}\bigg(-\frac{\sqrt{2E+2kX}}{2EX^{n-1}(n-1)}
-\frac{k(n-\frac{3}{2})}{E(n-1)}\int \frac{dX}{X^{n-1}\sqrt{2E+2kX}}\bigg).
\label{int03}
\end{eqnarray}

For integer values of $n$ the integrals (\ref{int02}) and (\ref{int03}) can be 
integrated explicitly by repeated use of these formulas. When $n$ is a
noninteger value the final term in the 
integrals (\ref{int02}) and (\ref{int03}) can be integrated in terms of 
logarithmic function or beta function \cite{Gradshteyn} 
(since the value of $n-1$ in the final integral lies between 0 to 1). We also
note here that for the special choices
$n=0$ or $-\frac{3}{2}$ (which corresponds to the case 
$\beta=\frac{\alpha^2}{9}$) and $n=-\frac{2}{3}$ or $-\frac{5}{6}$  (which
corresponds to the case $\beta=-\alpha^2$) respectively in  equations
(\ref{int02}) and (\ref{int03}), one can get the respective solutions reported
in the literature \cite{Painleve,Ince,Davis,
Kamke,Murphi,mahomed:1985,Duarte2,Bouquet,Sarleti,leach:1988a,Steeb,feix:1997,
Ibragimov1,Leach1,Chand,Chand1,Euler}. 

\subsection{Case 3: $\alpha^2<8\beta$}
\label{sec24}
Finaly, let us focus our attention in the regime $\alpha^2<8\beta$. In this 
parametric regime the Hamiltonian takes the form (vide equation (\ref{mlin16})) 
\begin{eqnarray}
H=\displaystyle{\frac{1}{2}\log[x^4 sec^2(\frac{\omega}{4} x^2p)]
-\frac{\alpha }{4}x^2p}.\label{com03a}
\end{eqnarray}
For the present case, let us choose the canonical transformation  
in the form
\begin{eqnarray}
&x=\frac{U}{P},\qquad\;\; p=\frac{P^2}{2},\label {com05aa}
\end{eqnarray}
and transform the Hamiltonian to  
\begin{eqnarray}
&H=\displaystyle{\frac{1}{2}\log[\frac{U^4}{P^4}sec^2(\frac{\omega}{8} U^2)]
-\frac{\alpha }{8}U^2} \equiv E.\label{com06a}
\end{eqnarray} 

Now making use of the canonical equations, $\dot{U}=-\frac{2}{P}$ and 
$\dot{P}=\frac{1}{4U}((\alpha-\omega\tan(\frac{\omega}{8} U^2))U^2-8)$, 
equation (\ref{com06a}) can be rewritten as
\begin{eqnarray}
E=\frac{1}{2}\log[\frac{\dot{U}^4U^4}{16}sec^2(\frac{\omega}{8} U^2)]
-\frac{\alpha }{8}U^2.
\label{int04}
\end{eqnarray}
Introducing now the transformation $V=\frac{U^2}{2}$ in (\ref{int04}) we arrive
at 
\begin{eqnarray}
E=\frac{1}{2}\log[\frac{\dot{V}^4}{16}sec^2(\frac{\omega}{4} V)]
-\frac{\alpha }{4}V.
\label{int04a}
\end{eqnarray}
Integrating the equation (\ref{int04a}) we get
\begin{eqnarray}
t-t_0=\frac{e^{-\frac{E}{2}}}{2}\int \sqrt{sec(\frac{\omega}{4} V)}
e^{\displaystyle{-\frac{\alpha }{8}V}} dV,
\label{int05}
\end{eqnarray}
where $t_0$ is the second integration constant. By substituting $W=e^V$ 
in (\ref{int05}) we obtain
\begin{eqnarray}
\fl \quad t-t_0=\frac{e^{-\frac{E}{2}}}{\sqrt{2}}
\int \frac{W^{q_1-1}dW}{\sqrt{1+W^{q_2}}}
=\frac{e^{-\frac{E}{2}}W^{q_1}}{\sqrt{2}q_1} F(\frac{1}{2},\frac{q_1}{q_2};
1+\frac{q_1}{q_2};-W^{q_2})
\label{int05a}
\end{eqnarray}
where $q_1=\frac{-\alpha+i\omega}{8},
\;q_2=\frac{i\omega}{2}$ and $F(\alpha,\beta;\gamma;z)$
is the hypergeometric function  \cite{Abramowitz,Gradshteyn}.

\section{Connection to two other nonlinear models}
\label{sec4}
In this section, we show that the MEE (\ref{eq01}) is intimately connected 
with two other 
well known nonlinear models, namely, the force-free Duffing oscillator type
equation and two dimensional Lotka-Volterra (LV) equation.
\subsection{Force-free Duffing oscillator type equation}
\label{sec31}
The MEE equation (\ref{eq01}) can be transformed to the following 
oscillator equation
\begin{eqnarray} 
w''+(\alpha w+\gamma)w'+\beta w^3+\frac{\alpha \gamma}{3}w^{2}
+\frac{2\gamma^2}{9}w=0,\;\;\;('=\frac{d}{d\tau})
 \label {eq08}
\end{eqnarray}
through the invertible point transformation $x = we^{\frac{\gamma}{3}\tau},
t = -\frac {3}{\gamma}e^{-\frac{\gamma}{3}\tau}$, where $\gamma$ is an arbitrary 
parameter. Equation (\ref{eq08}) includes 
force-free Duffing oscillator (in the specific case $\alpha=0$) 
\cite{Chand1,Bluman,Laks1} and quadratic oscillator with $\beta=0$
\cite{Chand1}. 

One can deduce the general solution of the equation (\ref{eq08})
from the general solution of MEE. For example, in the parametric choice 
$\beta=\frac{\alpha^2}{8}$ the general solution of the equation (\ref{eq08}) 
can be derived from (\ref{eq07}) in the form
\begin{eqnarray} 
w(\tau) =\frac{8}{\alpha}\mbox{erf}^{-1}(\hat{z})
e^{\frac{1}{2}\bigg(E-\frac{2}{3}\gamma\tau+2\mbox{erf}^{-1}(\hat{z})^2\bigg)},
\label{eq09}
\end{eqnarray}
where now $\displaystyle{\hat{z}=\frac{6e^{\frac{E}{2}}(\tau_0+e^{-\frac{\gamma}{3}\tau})}
{\gamma\sqrt{\pi}}}$ and $E$ and $\tau_0$ are arbitrary integration constants.
Similarly, one can fix the general solution for the equation (\ref{eq08}) in the
other parametric regimes also, that is, $\alpha^2>8\beta$ and $\alpha^2<8\beta$. 
\subsection{Two dimensional Lotka-Volterra equation}
\label{sec32}
Let us consider the two dimensional Lotka-Volterra equation of the form 
\begin{eqnarray} 
\dot {x}=x(a_1+a_2 x+a_3 y),\quad
\dot {y}=y(b_1+b_2x+b_3y),
\label{eq10}
\end{eqnarray}
where $a_i$'s and $b_i$'s are six real parameters. 
Equation (\ref{eq10}) models two species in competition in ecology and is
being analyzed for the past three decades or so in mathematical biology 
\cite{Murray,Cairo1,Cairo2}. Interestingly, equation (\ref{eq10}) can also be 
transformed to (\ref{eq08}) as follows. 
Rewriting the first equation in (\ref{eq10}) for the variable $y$ and substituting 
it into the second equation in equation (\ref{eq10}) we get the following second order
ODE for $x$, namely,
\begin{eqnarray} 
\fl \qquad\ddot{x}-(1+\frac{b_3}{a_3})\frac{\dot{x}^{2}}{x}+
((2a_2\frac{b_3}{a_3}-a_{2}-b_2)x+(2a_{1}\frac{b_3}
{a_3}-b_1))\dot{x}
\nonumber\\
\fl\qquad\quad+(b_2a_{2}-\frac{b_3}{a_3}a_2^{2})x^{3}
+(a_2 b_1+b_2a_{1}-2a_{1}a_2\frac{b_3}{a_3})x^{2}
+(a_1b_1-\frac{b_3}{a_3}a_{1}^{2})x=0.
\label {eq11}
\end{eqnarray}
Let us choose the parameters in (\ref{eq11}) in the form $b_3=-a_3$  
 and $b_1=a_1$ so that equation (\ref{eq11}) can be brought to the form
\begin{eqnarray} 
\fl\qquad\ddot{x}-\left((3a_2+b_2)x+3a_1\right)\dot{x}
+a_2(a_2+b_2)x^{3}+a_1(3a_2+b_2)x^{2}+2a_1^2x=0. \label {eq12}
\end{eqnarray}
The associated LV equation takes the form
\begin{eqnarray} 
\dot {x}=x(a_{1}+a_2x+a_3y),\quad
\dot {y}=y(a_{1}+b_2x-a_3y).
\label{eq13}
\end{eqnarray}
Now comparing the equations (\ref{eq08}) and (\ref{eq12}) we obtain 
$\alpha=-(3a_2+b_2)$, $\beta=a_2(a_2+b_2)$ and 
$\gamma=-3a_1$. Choosing $a_2=b_2$ ($\beta=\frac{\alpha^2}{8}$) the 
general 
solution for the LV equation (\ref{eq12}) can be obtained from (\ref{eq09}) 
and using this in the first equation in (\ref{eq13}) we arrive the general 
solution for the LV equation (\ref{eq13}) in the form  
\begin{eqnarray} 
\fl \qquad x(t) =-\frac{2}{a_{2}}\mbox{erf}^{-1}(z)
e^{\frac{1}{2}\bigg(E+2a_1t+2\mbox{erf}^{-1}(z)^2\bigg)},\quad
 y(t) = -\frac{e^{\frac{1}{2}\bigg(E+2a_1t+2\mbox{erf}^{-1}(z)^2\bigg)}}
{a_3\mbox{erf}^{-1}(z)},
\label{eq14}
\end{eqnarray}
where $\displaystyle{z=\frac{2e^{\frac{E}{2}}(t_0-e^{a_1t})}{a_1\sqrt{\pi}}}$ 
and $E$ and $t_0$ are arbitrary integration constants. 
Finally, we mention that the general solution 
for the equation (\ref{eq12}) for other parametric choices can be obtained 
from equation (\ref{eq08}) in 
a similar manner. Again (to our knowledge) the complete integrability of the LV system 
(\ref{eq13}) in these regimes is new to the literature. 
\section{Conclusions}
\label{sec5} 
In this paper, we have shown that the MEE (\ref{eq01}) is integrable either
explicitly or by quadratures for any value of $\alpha$ and $\beta$. 
We have also obtained the 
time independent Hamiltonians for
the equation (\ref{eq01}). We have transformed the Hamiltonians into simpler
forms, with appropriate canonical transformations, and deduced the general
solution by direct integration so that our approach 
helps to understand the dynamics of equation (\ref{eq01}) in phase-space
clearly. 
Further, we have demonstrated that the complete integrability of
equation (\ref{eq01}) also helps one to understand the dynamics of two other
nonlinear models, namely, the generalized oscillator equation and the two
dimensional LV equation. As a consequence the
solutions which  we have explored for the equation (\ref{eq01}) also provide
solutions for these  models as well. 

\section*{Acknowledgment}
The work of VKC is supported by CSIR in the form of a CSIR Senior Research
Fellowship.  The work of ML forms part of a Department of Science and 
Technology, Government of India sponsored research project and is supported by a
Department of Atomic Energy Raja Ramanna Fellowship.

\appendix

\section{Method of deriving integrals of motions}
\label{sec6}
In the following we briefly explain the generalized extended or modified
Prelle-Singer (PS) procedure for second order ODEs \cite{Chand,Chand1,Duarte} 
which is used to identify the integrals of motions (\ref{intm03})-(\ref{intm01}).

To begin with, let us rewrite equation~(\ref{eq01}) in the form 
\begin{eqnarray} 
\ddot{x}=-(\alpha x\dot{x}+\beta x^3)\equiv \phi(x,\dot{x}).
 \label{eq01a}
\end{eqnarray}
Further, we assume that the ODE (\ref{eq01a}) 
admits a first integral $I(t,x,\dot{x})=C,$ with $C$ constant on the 
solutions, so that the total differential becomes
\begin{eqnarray}  
dI={I_t}{dt}+{I_{x}}{dx}+{I_{\dot{x}}{d\dot{x}}}=0, 
\label{met3}  
\end{eqnarray}
where each subscript denotes partial differentiation with respect 
to that variable. Rewriting equation~(\ref{eq01a}) in the form 
$\phi dt-d\dot{x}=0$ and adding a null term 
$S(t,x,\dot{x})\dot{x}$ $ dt - S(t,x,\dot{x})dx$ to the latter, we obtain that on 
the solutions the 1-form
\begin{eqnarray}
(\phi +S\dot{x}) dt-Sdx-d\dot{x} = 0, \quad  \phi=-(\alpha x\dot{x}+\beta x^3).
\label{met6} 
\end{eqnarray}	
Hence, on the solutions, the 1-forms (\ref{met3}) and 
(\ref{met6}) must be proportional.  Multiplying (\ref{met6}) by the 
factor $ R(t,x,\dot{x})$ which acts as the integrating factors
for (\ref{met6}), we have on the solutions that 
\begin{eqnarray} 
dI=R(\phi+S\dot{x})dt-RSdx-Rd\dot{x}=0. 
\label{met7}
\end{eqnarray}
Comparing Eq.~(\ref{met3}) 
with (\ref{met7}) we have, on the solutions, the relations 
\begin{eqnarray} 
 I_{t}  = R(\phi+\dot{x}S),\quad 
 I_{x}  = -RS, \quad 
 I_{\dot{x}}  = -R.  
 \label{met8}
\end{eqnarray}
Then the compatibility conditions, 
$I_{tx}=I_{xt}$, $I_{t\dot{x}}=I_{{\dot{x}}t}$, $I_{x{\dot{x}}}=I_{{\dot{x}}x}$, 
between the Eqs.~(\ref{met8}), provide us
\begin{eqnarray}          
S_t+\dot{x}S_x+\phi S_{\dot{x}} &=& 
   -\phi_x+\phi_{\dot{x}}S+S^2,\label {lin02}\\
R_t+\dot{x}R_x+\phi R_{\dot{x}} & =&
-(\phi_{\dot{x}}+S)R,\label {lin03}\\
R_x-SR_{\dot{x}}-RS_{\dot{x}}  &= &0.
\qquad \qquad\qquad \;\;\;\label {lin04}
\end{eqnarray}

Solving equations~(\ref{lin02})-(\ref{lin04}) one can obtain expressions for $S$ and
$R$. It may be noted that any set of special solutions $(S,R)$ is sufficient for
our purpose. Once these forms are determined the integral of motion 
$I(t,x,\dot{x})$ can be deduced from the relation 
\begin{eqnarray}
 I= r_1
  -r_2 -\int \left[R+\frac{d}{d\dot{x}} \left(r_1-r_2\right)\right]d\dot{x},
  \label{met13}
\end{eqnarray}
where 
\begin{eqnarray} 
r_1 = \int R(\phi+\dot{x}S)dt,\quad
r_2 =\int (RS+\frac{d}{dx}r_1) dx. \nonumber
\end{eqnarray}
Equation~(\ref{met13}) can be derived straightforwardly by integrating the 
equation~(\ref{met8}).

As our motivation is to explore time independent integral of motion for the 
equation~(\ref{eq01a}) let us choose $I_t=0$.
In this case one can easily fix the null form $S$ from the first equation in
(\ref{met8}) as 
\begin{equation}
S = \frac{-\phi}{\dot{x}}=
\frac{(\alpha x\dot{x}+\beta x^3)}{\dot{x}}.
\label{mlin06}
\end{equation}
Substituting this form of $S$ into (\ref{lin03}) we get 
\begin{eqnarray}          
 \dot{x}R_x-(\alpha x\dot{x}+\beta x^3)R_{\dot{x}} 
 =-\frac{\beta x^3}{\dot{x}}R.
 \label {lin03a}
\end{eqnarray}

Equation~(\ref{lin03a}) is a first order linear partial differential equation
with variable coefficients.
As we noted earlier any particular solution is
sufficient to construct an integral of motion (along with the function $S$).
To seek
a particular solution for $R$ one can make a suitable ansatz instead of looking
for the general solution. We assume $R$ to be of the form,
\begin{equation}
R = \frac{\dot{x}}{(A(x)+B(x)\dot{x}+C(x)\dot{x}^2)^r},
\label{mlin08}
\end{equation} 
where $A,\;B$ and $C$ are functions of their arguments, and $r$ is a constant
which are all to be
determined. {\it We demand the above form of  ansatz (\ref{mlin08}), which is very
important to derive the Hamiltonian structure associated with the given 
equation}, due to the following
reason. To deduce the first integral $I$ we assume a rational form for $I$, 
that is,
$I=\frac{f(x,\dot{x})}{g(x,\dot{x})}$, where $f$ and $g$
are arbitrary functions of $x$ and $\dot{x}$ and are independent of $t$, from
which we get
$I_x=\frac{f_{x}g-fg_x}{g^2}$ and $I_{\dot{x}}=\frac{f_{\dot{x}}g-fg_{\dot{x}}}{g^2}$.
From (\ref{met8}) one can see that $R=I_{\dot{x}}=\frac{f_{\dot{x}}g-fg_{\dot{x}}}{g^2},
\;S=\frac{I_x}{I_{\dot{x}}}=\frac{f_xg-fg_x}{f_{\dot{x}}g-fg_{\dot{x}}}$ and
$RS=I_x$, so that the denominator of the function $S$ should be the numerator of the
function $R$. Since the denominater of $S$ is $\dot{x}$ (vide Eq.~(\ref{mlin06}))
we fixed the numerator of $R$ as $\dot{x}$. To seek a suitable function in the 
denominator initially one can consider an
arbitrary form $R=\frac{\dot{x}}{h(x,\dot{x})}$. However, it is difficult to
proceed with this choice of $h$. So let us assume that $h(x,\dot{x})$ is a 
function which is polynomial in
$\dot{x}$. To begin with let us consider the case where $h$ is quadratic in 
$\dot{x}$,
that is, $h=A(x)+B(x)\dot{x}+C(x)\dot{x}^2$, which is a generalized version of
the form considered in Ref. \cite{Chand1}, where only the linear form in
$\dot{x}$ was investigated (that is $C(x)=0$). Since $R$ is in rational form while
taking differentiation or integration the form of the denominator remains same 
but the  power of the denominator decreases or increases by a unit order from
that of the initial one. So instead of considering $h$ to be of the form
$h=A(x)+B(x)\dot{x}+C(x)\dot{x}^2$, one may consider a more general form
$h=(A(x)+B(x)\dot{x}+C(x)\dot{x}^2)^r$, where $r$ is a constant to be 
determined. The parameter $r$ plays an important role, as we see below. 

Substituting (\ref{mlin08}) into (\ref{lin03a}) and solving the resultant
equations, we arrive at the relation
\begin{eqnarray}
\fl \quad r\bigg[\dot{x}(A_x+B_x\dot{x}+C_x\dot{x}^2)
-(\alpha x\dot{x}+\beta x^3)(B+2C\dot{x})\bigg]
=-\alpha x(A+B\dot{x}+C\dot{x}^2).
\label{mlin09}
\end{eqnarray} 
Solving equation~(\ref{mlin09}), we can fix the forms of $A,\;B,\;C$ and $r$ and 
substituting them into equation~(\ref{mlin08}) we can get the integrating factor
$R$. Doing so, we find
\begin{eqnarray}
R=\left\{
\begin{array}{ll}
\displaystyle{\frac{\dot{x}}{(\dot{x}+\frac{(r-1)}{2r}\alpha x^2)^{r}}},
&\;\;\alpha^2\geq8\beta  \\
\displaystyle{\frac{\dot{x}}{(2\dot{x}^2+\alpha x^2\dot{x}+\beta x^4)}},
&\;\;\alpha^2 <8\beta \\
\displaystyle{\dot{x}}, & \;\; \alpha=0 
\end{array}\right.\label{lam107}
\end{eqnarray} 
where $r=\frac{\alpha}{4\beta}\bigg[\alpha\pm\sqrt{\alpha^2-8\beta}\bigg]$.
One can easily check the functions $S$ and $R$ given in (\ref{mlin06}) and 
(\ref{lam107}), respectively, satisfy (\ref{lin04}) also. Finally, substituting 
$R$ and $S$ into the form (\ref{met13}) for the integral we get the integrals of
motion (\ref{intm03})-(\ref{intm01}).

\end{document}